%% file: journalRevised.tex
\documentclass[9pt,technote,final]{IEEEtran}

\def\MODE{2}
\usepackage{lamperskiieeeLoc}
\author{Andrew~Lamperski\thanks{A. Lamperski is with the Department of Electrical and Computer Engineering, University
    of Minnesota, Minneapolis, MN, USA ({\tt alampers@umn.edu}).} and
John~C.~Doyle\thanks{J. C. Doyle is with Control and Dynamical Systems, California
  Institute of Technology, Pasadena, CA, USA ({\tt
    doyle@cds.caltech.edu})}}

\title{The $\mathcal{H}_2$ Control Problem for Quadratically Invariant Systems with Delays} 
\begin{document}
\maketitle
\input{abstract}
\input{intro}
\input{problem}
\input{stabilization}
\input{result}
\input{examples}
\input{conclusion}

\input{acknowledgement}
\bibliography{bibLoc}
\input{appendix}
\end{document}

%% file: abstract.tex
\begin{abstract}
This paper gives a new solution to the output feedback $\Htwo$
problem for quadratically invariant communication delay patterns. A characterization of all stabilizing controllers
satisfying the delay constraints is given and the decentralized
$\Htwo$ problem is cast as a convex model matching problem. The main
result shows that the model matching problem can be reduced to a
finite-dimensional quadratic program.  A recursive
state-space method
for computing the optimal controller based on vectorization is given.
\end{abstract}


%% file: intro.tex
\section{Introduction}

In decentralized control problems with delays,
inputs to a dynamic system are chosen by multiple controllers that pass their local measurements over a communication network with delays.
As a result, some controllers will have access to measurements before
others. This paper provides a new solution to the $\Htwo$ optimal
control problem, subject to quadratically invariant delay constraints,
based on the Youla 
parametrization and vectorization.  


\subsection{Contributions}

This paper solves the decentralized $\Htwo$ problem for a class of
delay patterns arising from strongly-connected communication
networks. The delay constraints are assumed to be quadratically invariant, which implies that the optimal control problem is convex. 
The main contribution of the paper is a reformulation of the
decentralized $\Htwo$ problem for such delay patterns as a
finite-dimensional quadratic program.
This quadratic program, in turn,
can be solved as a finite-horizon linear quadratic regulator problem. 

To derive the quadratic program, a Youla parametrization
framework  developed for sparsity problems, \cite{sabaunecessary2011},
is adapted to
communication delay patterns. The parametrization is then used to
characterize all stabilizing controllers that satisfy
a given delay pattern. It is then shown that for a doubly-coprime
factorization based on the centralized LQG controller, the
corresponding model matching problem reduces to a quadratic
program. Finally, the quadratic program is cast as a finite-horizon linear
quadratic regulator problem using vectorization. 

\subsection{Related Work}

This paper focuses on the $\Htwo$ problem subject to a general class
of quadratically invariant delay constraints. 
Existing approaches to
this problem are based on  
vectorization \cite{rotkowitzcharacterization2006} and linear
matrix inequalities (LMIs) 
\cite{rantzerseparation2006,gattamigeneralized2006}. In those works,
the decentralized problems are reduced to centralized control problems with
state dimensions that grow with the size of the delay. This paper, on
the other hand, shows that the solution can be computed in terms of
the classical centralized solution and a quadratic program. This
quadratic program, in turn, may be interpreted as a finite-horizon
control problem with fixed dimension but horizon growing the with the
size of the delay. 

For specific delay patterns, dynamic programming techniques exist to solve output feedback decentralized LQG
problems
\cite{sandellsolution1974,kurtaranlinearquadraticgaussian1974,yoshikawadynamic1975,feyzmahdaviandistributed2012}. These
delay patterns all satisfy the condition known as partial nestedness
\cite{hoteam1972}, which is closely related to quadratic invariance
\cite{rotkowitzinformation2008}, and guarantees that the optimal
policies are linear functions of the measurements. For more general
partially nested
delay constraints, dynamic programming methods for linear quadratic
state feedback are known,
\cite{lamperskidynamic2012,lamperskioptimal2012}. 
New results have identified sufficient statistics for dynamic
programming in decentralized problems, without partial-nestedness assumptions,
\cite{nayyardecentralized2013,mahajansufficient2014}, but they do not
provide  solutions to the corresponding LQG problems. 

This paper uses an operator theoretic approach to solve decentralized
$\Htwo$ problems with delays. 
 It is an extension of
\cite{lamperskioutput2013}, which uses 
spectral factorization to derive a similar quadratic
program. Many of the calculations
are modified from spectral factorization methods for sparsity
constraints such as
\cite{swigartexplicit2010a,shahh2optimal2010,lessardoptimal2012}. Another
operator theoretic approach, based on loop-shifting
\cite{mirkindeadtime2011}, has also been developed for special
quadratically invariant delay patterns \cite{kristalnydecentralized2012}.

\subsection{Overview}

The paper is structured as follows. Section \ref{sec:problem}
defines the general problem studied in this paper, the decentralized
$\Htwo$ problem with a strongly-connected delay pattern. Section
\ref{sec:stabilization} gives a parametrization of all stabilizing
controllers that satisfy a given delay pattern, and presents the
corresponding model matching problem. In Section \ref{sec:results}, the decentralized $\Htwo$
problem is reduced to a quadratic program, and this program is solved
by vectorization.
Numerical
results are given in Section \ref{sec:examples} and finally, conclusions are
given in \ref{sec:conclusion}.


%% file: problem.tex
\section{Problem}
\label{sec:problem}

This section introduces the basic notation and the control 
problem of interest. Subsection \ref{sec:communication} describes how delayed information sharing patterns can be cast in the
framework of this paper.  

\subsection{Preliminaries}
Let $\D = \{z\in \C:|z|<1\}$ be the unit disc of complex numbers and
let $\overline{\D}$ be its closure. 
Let $\Htwo$ and $\Hinf$ denote the
Hardy spaces of matrix-valued functions that are analytic on
$(\C\cup\{ \infty\})\setminus \overline{\D}$. 

Let $\RRp$ denote the space of proper real rational transfer
matrices. Furthermore, denote $\RRp\cap \Htwo$ and $\RRp\cap \Hinf$ by
$\RR\Htwo$ and $\RR\Hinf$, respectively. Note that
$\RR\Htwo=\RR\Hinf$, since both correspond to transfer matrices with
no poles outside of $\D$. 

A function $G(z)\in\Htwo$ has a power series expansion given by  
$
G(z) = \sum_{i=0}^{\infty}\frac{1}{z^i}G_i.
$
Furthermore, $\Htwo$ is a Hilbert space with inner product defined by 
\if\MODE1
\[
\langle G,H\rangle =
\lim_{r\downarrow 1}
 \frac{1}{2\pi}\int_{-\pi}^{\pi}\Tr\left(
G\left(re^{j\theta}\right)H\left(re^{j\theta}\right)^{\sim} 
\right)
 d\theta 
= \sum_{i=0}^{\infty} \Tr\left(
G_iH_i^*
\right),
\]
\fi
\if\MODE2
\begin{eqnarray*}
\langle G,H\rangle &=&
\lim_{r\downarrow 1}
 \frac{1}{2\pi}\int_{-\pi}^{\pi}\Tr\left(
G\left(re^{j\theta}\right)H\left(re^{j\theta}\right)^{\sim} 
\right)
 d\theta \\
&=& \sum_{i=0}^{\infty} \Tr\left(
G_iH_i^*
\right),
\end{eqnarray*}
\fi
where the second equality follows from Parseval's identity. 

Define the conjugate of $G$ by 
$
G(z)^{\sim} = \sum_{i=0}^{\infty}z^iG_i^*
$.
For $G = \left[\begin{array}{c|c}
A & B \\
\hline 
C & D
\end{array}
\right]
\in\RRp$ , the conjugate is given by 
\[
\left(C(zI-A)^{-1}B+D\right)^{\sim} = 
B^\tp\left(
\frac{1}{z}I-A^\tp
\right)^{-1}C^\tp+D^\tp.
\]

If $\mathcal{M}$ is a subspace of $\Htwo$, denote the orthogonal
projection onto $\mathcal{M}$ by $\P_{\mathcal{M}}$.


\subsection{Formulation}

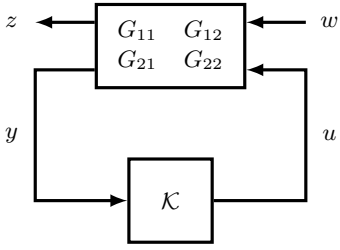
\begin{figure}
\centering
\begin{tikzpicture}[very thick,node distance=6.5em,>=latex]
  \tikzstyle{block}=[rectangle,draw,minimum height=3.5em,minimum width=3.5em]

  \def\blockdist{2.5em}
  \def\inputoffset{1em}
  \def\labeloffset{3.5em}

  \node[block] (P) 
{
$
\begin{array}{cc}
G_{11} & G_{12}\\
G_{21} & G_{22}
\end{array}
$
};

\node[block, below of=P] (K) {$
\K
$};

\draw[->] let \p1 = (P.east),
\p2 = (K.east) in
(\p2) -- ({\x1 + \blockdist},\y2)
-- ({\x1+\blockdist},{\y1-\inputoffset})
-- (\x1,{\y1-\inputoffset});

\draw[->] let \p1 = (P.west),
\p2 = (K.west) in 
(\x1,{\y1-\inputoffset})--({\x1-\blockdist},{\y1-\inputoffset})
--  ({\x1-\blockdist},\y2)
--(\p2);

\draw[->] let \p1 = (P.east) in
({\x1+\blockdist},{\y1+\inputoffset}) -- 
(\x1,{\y1+\inputoffset});

\draw[->] let \p1 = (P.west) in 
(\x1,{\y1+\inputoffset}) -- 
({\x1-\blockdist},{\y1+\inputoffset});

\path let \p1 = (P.west) in 
({\x1-\labeloffset},{\y1+\inputoffset}) node (z) {$z$};

\path let \p1 = (P.east) in 
(\p1)+(\labeloffset,\inputoffset) node (w) {$w$};

\path let \p1 = (P.east), \p2 = (K.east) in
({\x1+\labeloffset},{(\y1-\inputoffset+\y2)/2}) node {$u$};

\path let \p1 = (P.west), \p2 = (K.east) in 
({\x1-\labeloffset},{(\y1-\inputoffset+\y2)/2}) node {$y$};
\end{tikzpicture}
\caption{\label{fig:loop}
The basic feedback loop.
}
\end{figure}

This subsection introduces the generic problem of interest. Let $G$ be a discrete-time plant given by 
\[
G = 
\left[
\begin{array}{c|cc}
A & B_1 & B_2 \\
\hline
C_1 & 0 & D_{12} \\
C_2 & D_{21} & 0
\end{array}
\right] = 
\begin{bmatrix}
G_{11} & G_{12} \\
G_{21} & G_{22}
\end{bmatrix},
\]
with inputs of dimension $p_1$, $p_2$ and outputs of dimension $q_1$, $q_2$. Let $\K$ be a feedback controller connected to $G$ as in Figure~\ref{fig:loop}. 

For the existence of solutions of the appropriate Riccati equations,
as well as simplicity of formulas, 
assume that 
\begin{itemize}
\item $(A,B_1,C_1)$ is stabilizable and detectable,
\item $(A,B_2,C_2)$ is stabilizable and detectable,
\item $D_{12}^T\begin{bmatrix}C_1 & D_{12}\end{bmatrix} = 
\begin{bmatrix} 0 & I \end{bmatrix}$,
\item $D_{21}\begin{bmatrix}
B_1^T & D_{21}^T
\end{bmatrix} = \begin{bmatrix} 0 & I \end{bmatrix}$.
\end{itemize}

For $N\ge 1$, define the spaces of proper and strictly proper finite impulse
response (FIR) transfer matrices by $\mathcal{X}_{\mathrm{p}} = \bigoplus_{i=0}^N\frac{1}{z^i} \C^{p_2\times q_2}$ and
 $\mathcal{X} =
\bigoplus_{i=1}^N\frac{1}{z^i}\C^{p_2\times q_2}$, respectively. Denote the corresponding spaces of real FIR transfer matrices by 
$\RR\mathcal{X}_{\mathrm{p}} =  \bigoplus_{i=0}^N\frac{1}{z^i}\R^{p_2\times q_2}$ 
and
$\RR\mathcal{X} =  \bigoplus_{i=1}^N\frac{1}{z^i}\R^{p_2\times q_2}$.
Note that $\Htwo$ and 
$\frac{1}{z}\Htwo$ 
can thus be decomposed into orthogonal subspaces as 
\begin{equation}\label{eq:HtwoSplit}
\Htwo = \mathcal{X}_{\mathrm{p}} \oplus \frac{1}{z^{N+1}} \Htwo
\qquad\textrm{and}\qquad
\frac{1}{z}\Htwo = \mathcal{X} \oplus \frac{1}{z^{N+1}} \Htwo. 
\end{equation}
 
Let $\mathcal{S}\subset\frac{1}{z}\mathcal{R}_{\mathrm{p}}$
be a subspace of 
the form 
\begin{equation}\label{eq:subspace}
\mathcal{S} =
\mathcal{Y}\oplus\frac{1}{z^{N+1}}\mathcal{R}_{\mathrm{p}},
\quad \textrm{where} \quad
\mathcal{Y} = \bigoplus \frac{1}{z^i} \mathcal{Y}_i,
\end{equation}
and $\mathcal{Y}_i\subset \R^{p_2\times q_2}$ defines a sparsity
pattern over matrices. Delay patterns satisfying the decomposition in
\eqref{eq:subspace} will be called {\it strongly connected},
since delay patterns arising from strongly-connected communication
networks always have this form. (See subsection
\ref{sec:communication}.) 

The set $\mathcal{S}$ is assumed to be {\it quadratically invariant}
with respect to $G_{22}$, which means that for all $\mathcal{K}\in
\mathcal{S}$, $\mathcal{K}G_{22}\mathcal{K}\in\mathcal{S}$. The key
property of quadratic invariance is that $\K\in \mathcal{S}$ if and
only if $\K(I-G_{22}\K)^{-1}\in\mathcal{S}$
\cite{rotkowitzcharacterization2006}. 

The decentralized $\Htwo$ problem studied in this paper is given by
\begin{equation}\label{eq:standardProb}
\begin{array}{c}
\min_{\mathcal{K}}
 \|G_{11}+G_{12}\mathcal{K}(I-G_{22}\mathcal{K})^{-1}G_{21}\|_{\Htwo}^2 \\
\textrm{s.t. } \mathcal{K}\in \mathcal{S}.
\end{array}
\end{equation}

The quadratic invariance assumption guarantees that the corresponding
model matching problem is convex
\cite{rotkowitzcharacterization2006}. 
Reduction to model matching is
discussed in Section \ref{sec:modelMatching}.

The decomposition of $\mathcal{S}$ in \eqref{eq:subspace} is crucial
for the results of this paper. The property that $\frac{1}{z^{N+1}}\RRp\subset
\mathcal{S}$ implies that every measurement is available to all
controller subsystems within $N$ time
steps. Concrete examples of delay patterns of this form are described
in the next subsection. 

For technical simplicity, controllers in this paper are assumed to be
strictly proper (that is, in
$\frac{1}{z}\mathcal{R}_{\mathrm{p}}$). The results in this paper can
be extended to non-strictly proper controllers but more complicated
formulas would result. 

\subsection{Communication Delay Patterns}

\label{sec:communication}

\begin{figure}
\centering
\begin{tikzpicture}[thick,node distance=2.0cm,>=latex]
	\def\fs{\small}  
	\tikzstyle{block}=[circle,draw,fill=black!6,minimum height=2.1em]
	\tikzstyle{loopstyle}=[loop,looseness=6,dashed]
	\node [block](N1){$1$};
	\node [block, right of=N1](N2){$2$};
	\draw [->] (N1) edge [out=30, in=150]  node[above]{\fs$N$} (N2);
	\draw [<-] (N1) edge [out=-30,in=-150] node[below]{\fs$N$} (N2);
        \draw [->] (N1) edge [out=-150,in=150,loopstyle]
        node[left]{\fs$1$} (N1);
        \draw [->] (N2) edge [out=-30,in=30,loopstyle]
        node[right]{\fs$1$} (N2);
\end{tikzpicture}
\caption{\label{fig:Nstep}
The strictly proper $N$-step delay information pattern can be visualized as a
  two-node graph. The delay-$1$ self-loops specify computational
  delays of $1$ at each node, while the delay-$N$ edges specify
  communication delays. Self-loops are drawn as dashed arrows to
  distinguish them as denoting computational delays.} 
\end{figure}
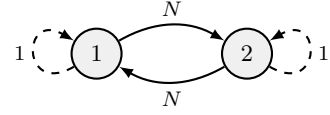

This subsection
will discuss how \eqref{eq:subspace} can be used to model delay
patterns that arise
from strongly connected graphs. 
As an example, consider an $N$-step delayed information pattern, represented by \eqref{eq:subspace}
with $\mathcal{Y}$ corresponding to 
block diagonal FIR matrices
\[
\mathcal{Y} = \bigoplus_{i=1}^{N} \frac{1}{z^i} 
\begin{bmatrix}
\R^{p_{21}\times q_{21}} & 0 \\
0 & \R^{p_{22} \times q_{22}}
\end{bmatrix}.
\] 
 The corresponding graph is given in Figure~\ref{fig:Nstep}. It was shown in \cite{varaiyadelayed1978} that the separation principle conjectured in \cite{witsenhausenseparation1971} fails when $N\ge 2$, and appropriate sufficient statistics were given in \cite{nayyaroptimal2011,mahajansufficient2014}. The special case of $N=1$ was solved explicitly in
\cite{sandellsolution1974,kurtaranlinearquadraticgaussian1974,yoshikawadynamic1975}.

More generally, assume that communication between the controller
subsystems is specified by a strongly-connected graph 
$(V,E)$ with self-loops at each node. Computational delays are
specified by positive integers
on the self-loops, while communication delays are represented by
non-negative integers on the edges between distinct nodes. Requiring
positive computational delays ensures that the controller is strictly
proper. 

A constraint space of the form \eqref{eq:subspace} can be constructed as
follows. For nodes $i$ and $j$ let $c_i$ be the computational delay at
node $i$ and let $\tilde{d}_{ij}$ be the sum of communication delays
along the directed path with shortest aggregate delay. Let the {\it
  delay matrix}, $d$, be the matrix with entries
$
d_{ij} = c_i + \tilde{d}_{ij}.
$
In the $N$-step delay example, the delay matrix is given by
$
d = \begin{bmatrix}
1 & N+1 \\
N+1 & 1
\end{bmatrix}.
$

Let $N=\max\{d_{ij}:i,j\in V\}-1$.\footnote{Using this convention, all
  measurements, $y_j(t)$, are available to all controllers by time $t+N+1$.
}
 The
corresponding constraint space is defined by 
\[
\mathcal{S} = \begin{bmatrix}
\frac{1}{z^{d_{11}}} \RRp & \cdots & \frac{1}{z^{d_{1|V|}}} \RRp \\
\vdots & & \vdots \\
\frac{1}{z^{d_{|V|1}}}\RRp & \cdots & \frac{1}{z^{|V||V|}}\RRp
\end{bmatrix}.
\]
Thus, the $\mathcal{S}$ can be decomposed as in 
\eqref{eq:subspace} by defining
\if\MODE1
\[
\mathcal{Y} = \bigoplus_{k=1}^N\frac{1}{z^k}
\begin{bmatrix}
\mathcal{Y}_k^{11} & \cdots  & \mathcal{Y}_k^{1|V|} \\
\vdots & & \vdots \\
\mathcal{Y}_k^{|V|1} & \cdots & \mathcal{Y}_k^{|V||V|}
\end{bmatrix}
\qquad\textrm{where}\qquad
\mathcal{Y}_k^{ij} =
\left\{
\begin{array}{cc}
\R^{p_{2i}\times q_{2j}} & \textrm{if } d_{ij} \le k \\
0 & \textrm{if } d_{ij} > k.
\end{array}
\right.
\]
\fi
\if\MODE2
\[
\mathcal{Y} = \bigoplus_{k=1}^N\frac{1}{z^k}
\begin{bmatrix}
\mathcal{Y}_k^{11} & \cdots  & \mathcal{Y}_k^{1|V|} \\
\vdots & & \vdots \\
\mathcal{Y}_k^{|V|1} & \cdots & \mathcal{Y}_k^{|V||V|}
\end{bmatrix}
\]
where 
$
\mathcal{Y}_k^{ij} =
\left\{
\begin{array}{cc}
\R^{p_{2i}\times q_{2j}} & \textrm{if } d_{ij} \le k \\
0 & \textrm{if } d_{ij} > k.
\end{array}
\right.
$
\fi

Let the blocks of $G_{22}$ satisfy $(G_{22})_{ij}\in \frac{1}{z^{p_{ij}}} \RR_p$. It was shown in \cite{rotkowitztractable2005} that $\mathcal{S}$ defined above is quadratically invariant with respect to $G_{22}$ if and only if 
\[
d_{ki} + p_{ij} + d_{jl} \ge d_{kl} \textrm{ for all } i,j,k,l.
\]
This constraint guarantees that signals travel through the controller network at least as fast as through the plant.

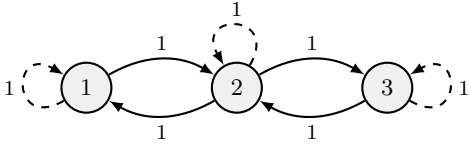
\begin{figure}
\centering
\begin{tikzpicture}[thick,node distance=2.0cm,>=latex]
	\def\fs{\small}  
	\tikzstyle{block}=[circle,draw,fill=black!6,minimum height=2.1em]
	\tikzstyle{loopstyle}=[loop,looseness=6,dashed]
	\node [block](N1){$1$};
	\node [block, right of=N1](N2){$2$};
	\node [block, right of=N2](N3){$3$};
	\draw [->] (N1) edge [out=30, in=150]  node[above]{\fs$1$} (N2);
	\draw [<-] (N1) edge [out=-30,in=-150] node[below]{\fs$1$}
        (N2);
        \draw [->] (N2) edge [out=30, in=150]  node[above]{\fs$1$} (N3);
	\draw [<-] (N2) edge [out=-30,in=-150] node[below]{\fs$1$} (N3);
        \draw [->] (N1) edge [out=-150,in=150,loopstyle]
        node[left]{\fs$1$} (N1); 
        \draw [->] (N2) edge [out=60,in=120,loopstyle]
        node[above]{\fs$1$} (N2);
        \draw [->] (N3) edge [out=-30,in=30,loopstyle]
        node[right]{\fs$1$} (N3);
\end{tikzpicture}
\caption{\label{fig:chain} The network graph for the
  three-player chain. The self-loops specify computational delays, while solid edges specify communication delays.}
\end{figure}

As another example, consider the 
strictly proper version of the three-player chain problem discussed in
\cite{lamperskistructure2011,lamperskioutput2013}. The graph
describing the delays is given in Figure~\ref{fig:chain}, leading to a delay matrix and FIR constraint space
\begin{equation}
\label{eq:chain}
d = \begin{bmatrix}
1 & 2 & 3 \\
2 & 1 & 2 \\
3 & 2 & 1
\end{bmatrix} \quad \textrm{and} \quad
\mathcal{Y} = \frac{1}{z}
\begin{bmatrix}
* & 0 & 0 \\
0 & * & 0 \\
0 & 0 & *
\end{bmatrix}\oplus
\frac{1}{z^2}
\begin{bmatrix}
* & * & 0 \\
* & * & * \\
0 & * & *
\end{bmatrix},
\end{equation}
respectively. For compactness, $*$ is used to denote a space of appropriately
sized real matrices. 

%% file: stabilization.tex
\section{Decentralized Stabilization}\label{sec:stabilization}

This section parametrizes the set of controllers $\K \in\mathcal{S}$
which internally stabilize the plant $G$. The parametrization naturally
leads to a convex model matching formulation of $\Htwo$ problem. In
analogy with results on  
sparse transfer matrices \cite{sabaunecessary2011}, the
parametrization is based on quadratic invariance and the classical
Youla parametrization. 

\subsection{All Stabilizing Decentralized Controllers}

A collection of stable transfer matrices, $\hat M$, $\hat N$, $\hat X$, $\hat Y$, $\tilde M$, $\tilde N$, $\tilde X$, and $\tilde Y$, defines a {\it doubly-coprime factorization} of $G_{22}$ if $G_{22} = \hat N \hat M^{-1} = \tilde M^{-1} \tilde N$ and 
\begin{equation}
\label{bezout}
\begin{bmatrix}
\tilde{X} & -\tilde{Y} \\
-\tilde{N} & \tilde{M}
\end{bmatrix}\begin{bmatrix}
\hat{M} & \hat{Y} \\
\hat{N} & \hat{X}
\end{bmatrix} = I.
\end{equation}
As long as $(A,B_2,C_2)$ is stabilizable and detectable, there are numerous ways to construct a
doubly coprime factorization of $G_{22}$. 

The following theorem is well known \cite{dullerudcourse2000}. 

\gap

\begin{theorem}\label{thmYoula}
{\it
Assume that $G_{22}$ has a double doubly-coprime
factorization of the form in \eqref{bezout}. 
A controller $\K\in\RRp$ internally stabilizes $G$ if and only if
there is a transfer matrix $Q\in \RR\Hinf$ such that 
\begin{equation}
\label{youla}
\K = (\hat{Y}-\Hat{M}Q)(\hat{X}-\hat{N}Q)^{-1}
=(\tilde{X}-Q\tilde{N})^{-1}(\tilde{Y}-Q\tilde{M}).
\end{equation}
}
\end{theorem}

\gap

From \cite{rotkowitzcharacterization2006}, if $G_{22}$ is
quadratically invariant under $\mathcal{S}$, then $\K\in\mathcal{S}$ if and
only if  $\K(I-G_{22}\K)^{-1}\in \mathcal{S}$. As in
\cite{sabaunecessary2011}, a straightforward calculation shows that 
\begin{equation}
\label{linearLFT}
\K(I-G_{22}\K)^{-1} = (\hat Y - \hat M Q)\tilde M,
\end{equation}
and thus 
$
\K\in\mathcal{S} \iff (\hat Y - \hat M Q)\tilde
M\in \mathcal{S}.
$

Based on \eqref{eq:HtwoSplit}, $Q\in \RR\Htwo=\RR\Hinf$ can be decomposed uniquely as $Q=U+V$ with $U\in\frac{1}{z^{N+1}}\RR\Htwo$ and $V\in\mathcal{X}_{\mathrm{p}}$. Recalling 
\eqref{eq:subspace} and noting that $\hat M U \tilde M \in \frac{1}{z^{N+1}}\Htwo$ implies that 
\if\MODE1
\[
(\hat Y - \hat M Q)\tilde M \in \mathcal{S} \iff \P_{\mathcal{X}_{\mathrm{p}}}((\hat Y - \hat M Q)\tilde M)
\in\mathcal{Y} \iff \P_{\mathcal{X}_{\mathrm{p}}}((\hat Y - \hat M V)\tilde M) \in \mathcal{Y}.
\]
\fi
\if\MODE2
\begin{align*}
(\hat Y - \hat M Q)\tilde M \in \mathcal{S} 
&\iff \P_{\mathcal{X}_{\mathrm{p}}}((\hat Y - \hat M Q)\tilde M)
\in\mathcal{Y} 
\\
&\iff \P_{\mathcal{X}_{\mathrm{p}}}((\hat Y - \hat M V)\tilde M) \in \mathcal{Y}.
\end{align*}
\fi
Thus, the following characterization of all stabilizing decentralized controllers holds. 

\begin{theorem}\label{thmYoulaDecentralized}
{\it
A controller $\K\in\mathcal{S}$ internally stabilizes $G_{22}$ if and
only if there are transfer matrices $U\in\frac{1}{z^{N+1}}\RR\Hinf$
and $V\in \RR\mathcal{X}_{\mathrm{p}}$ 
 such that 
$
\K = (\hat Y - \hat M(U+V))(\hat X - \hat N(U+V))^{-1}
$
and 
\begin{equation}
\label{youlaFIR}
\P_{\mathcal{X}_{\mathrm{p}}}\left(
(\hat Y - \hat M V)\tilde M
\right) \in \mathcal{Y}.
\end{equation}
}
\end{theorem}




\gap

Note that \eqref{youlaFIR} reduces to a finite-dimensional linear
constraint on the FIR term, $V\in \RR\mathcal{X}_{\mathrm{p}}$. The other term, $U$, is delayed, but otherwise unconstrained. 

\subsection{Model Matching}\label{sec:modelMatching}

Given a doubly-coprime factorization, \eqref{linearLFT} implies that the closed-loop transfer matrix is given by
\[
G_{11}+G_{12}\K(I-G_{22}\K)^{-1}G_{21} = P_{11}+P_{12}QP_{21},
\]
where
\if\MODE1
\begin{equation}\label{modelMatchingMatrices}
P_{11} = G_{11}+G_{12}\hat Y \tilde M G_{21},\qquad
P_{12} = -G_{12}\hat M,\qquad 
P_{21} = \tilde M G_{21}. 
\end{equation}
\fi
\if\MODE2
\begin{align}
\nonumber
P_{11} &= G_{11}+G_{12}\hat Y \tilde M G_{21} \\
\label{modelMatchingMatrices}
P_{12} &= -G_{12}\hat M \\
\nonumber
P_{21} &= \tilde M G_{21}. 
\end{align}
\fi
Using the decomposition $Q=U+V$, with $V\in\RR\mathcal{X}_{\mathrm{p}}$ and
$U\in\frac{1}{z^{N+1}}\RR\Htwo$, the decentralized $\Htwo$ problem,
\eqref{eq:standardProb}, is equivalent to the following {\it model matching
problem}:
\if\MODE1
\begin{equation}\label{eq:modelMatching}
\begin{array}{cc}
\min_{U,V} & \|P_{11}+P_{12}(U+V)P_{21}\|_{\Htwo}^2 \\
\textrm{s.t.} & U\in\frac{1}{z^{N+1}}\RR\Htwo,\quad V\in
\RR\mathcal{X}_{\mathrm{p}},\quad 
\P_{\mathcal{X}_{\mathrm{p}}}\left(
(\hat Y - \hat M V)\tilde M
\right) \in \mathcal{Y}.
\end{array}
\end{equation}
\fi
\if\MODE2
\begin{equation}\label{eq:modelMatching}
\begin{array}{cc}
\min_{U,V} & \|P_{11}+P_{12}(U+V)P_{21}\|_{\Htwo}^2 \\
\textrm{s.t.} & U\in\frac{1}{z^{N+1}}\RR\Htwo,\quad V\in
\RR\mathcal{X}_{\mathrm{p}} \\
& \P_{\mathcal{X}}\left(
(\hat Y - \hat M V)\tilde M
\right) \in \mathcal{Y}.
\end{array}
\end{equation}
\fi


%% file: result.tex
\section{Results}\label{sec:results}

This section  gives the main result of the paper, a reduction of the decentralized control problem,
\eqref{eq:standardProb}, to a quadratic program. 
A vectorization method for
computing the optimal solution is also given.  

\subsection{Quadratic Programming Formulation}

In the previous section, it was shown that
the decentralized feedback problem is equivalent to a model matching
problem, \eqref{eq:modelMatching}. 
It will be shown 
that for a special doubly-coprime factorization, the
model matching problem reduces to a quadratic program. 

Let $X$ and $Y$ be the stabilizing solutions of the Riccati equations associated with the linear quadratic regulator and Kalman filter, respectively:
\begin{align}
\label{controlRiccati}
X& = C_1^\tp C_1 + A^\tp X A - A^\tp XB_2 (I+B_2^\tp XB_2)^{-1}  B_2^\tp X A \\
\label{filteringRiccati}
Y& = B_1B_1^\tp +AYA^\tp - AYC_2^\tp (I+C_2YC_2^\tp)^{-1} C_2 Y A^\tp.
\end{align}
Define $\Omega = I+B_2^\tp X B_2$ and $\Psi = I+C_2 Y C_2^\tp$. The corresponding gains are given by 
\if\MODE1
\begin{equation}
\label{gains}
K= -\Omega^{-1} B_2^\tp X A,\qquad
L%
= -AYC_2^\tp\Psi^{-1}.
\end{equation}
\fi
\if\MODE2
\begin{align}
\label{controlGain}
K&= -\Omega^{-1} B_2^\tp X A\\
\label{filterGain}
L& = -AYC_2^\tp\Psi^{-1}. 
\end{align}
\fi
Furthermore, $A+B_2K$ and $A+LC_2$ are stable. 

It is well known (e.g. \cite{dullerudcourse2000}) that a
doubly-coprime factorization of $G_{22}$ is given by  

\if\MODE1
\begin{equation}
\label{ssDCF}
\begin{bmatrix}
\hat M & \hat Y \\
\hat N & \hat X
\end{bmatrix} =
\left[
\begin{array}{c|cc}
A+B_2K & B_2 & -L \\
\hline
K & I & 0 \\
C_2 & 0 & I
\end{array}
\right],\quad
\begin{bmatrix}
\tilde X & -\tilde Y \\
-\tilde N & \tilde M
\end{bmatrix}=
\left[
\begin{array}{c|cc}
A+LC_2 & B_2 & -L \\
\hline
-K & I & 0 \\
-C_2 & 0 & I
\end{array}
\right].
\end{equation}
\fi
\if\MODE2
\begin{equation}
\label{ssDCF}
\begin{aligned}
\begin{bmatrix}
\hat M & \hat Y \\
\hat N & \hat X
\end{bmatrix} &=
\left[
\begin{array}{c|cc}
A+B_2K & B_2 & -L \\
\hline
K & I & 0 \\
C_2 & 0 & I
\end{array}
\right],
\\
\begin{bmatrix}
\tilde X & -\tilde Y \\
-\tilde N & \tilde M
\end{bmatrix}&=
\left[
\begin{array}{c|cc}
A+LC_2 & B_2 & -L \\
\hline
-K & I & 0 \\
-C_2 & 0 & I
\end{array}
\right].
\end{aligned}
\end{equation}
\fi

The following theorem is the main result of the paper.

\begin{theorem}\label{thm:modelMatchQP}
{\it
Consider the doubly-coprime factorization of $G_{22}$ defined by \eqref{ssDCF}. 
The optimal solution to the decentralized $\Htwo$ problem defined by \eqref{eq:standardProb} is given by 
\[
\K^{*} = (\hat Y - \hat MV^*)(\hat X-\hat NV^*)^{-1},
\]
where $V^*$ is the unique optimal solution to the quadratic program
\begin{equation}\label{simplifiedModelMatching}
\begin{array}{cc}
\min_{V\in \RR\mathcal{X}} &\|\Omega^{1/2} V \Psi^{1/2}\|_{\Htwo}^2 \\
\textrm{s.t.}& \P_{\mathcal{X}}\left(
(\hat Y - \hat M V)\tilde M
\right) \in \mathcal{Y}.
\end{array}
\end{equation}
Furthermore, the optimal cost is given by 
$
\|P_{11}\|_{\Htwo}^2+\|\Omega^{1/2} V^* \Psi^{1/2}\|_{\Htwo}^2.
$
}
\end{theorem}

\begin{IEEEproof} 
For the doubly-coprime factorization given by \eqref{ssDCF}
the model matching matrices, \eqref{modelMatchingMatrices}, have state space realizations given by
\if\MODE1
\begin{equation}
\label{modelMatchMatrices}
\begin{array}{c}
P_{11} = \left[
\begin{array}{cc|c}
A+B_2 K & -B_2K & B_1 \\
0 & A+LC_2 & B_1 +LD_{21}\\ 
\hline 
C_1+D_{12}K & -D_{12}K & 0 
\end{array}
\right]
\\
P_{12} = -\left[
\begin{array}{c|c}
A+B_2K & B_2 \\
\hline
C_1 +D_{12}K & D_{12}
\end{array}
\right],\qquad
P_{21} = \left[
\begin{array}{c|c}
A+LC_2 & B_1+LD_{21} \\
\hline
C_2 & D_{21}
\end{array}
\right].
\end{array}
\end{equation}
\fi
\if\MODE2
\begin{align}
\nonumber
P_{11} &= \left[
\begin{array}{cc|c}
A+B_2 K & -B_2K & B_1 \\
0 & A+LC_2 & B_1 +LD_{21}\\ 
\hline 
C_1+D_{12}K & -D_{12}K & 0 
\end{array}
\right]
\\
\label{modelMatchMatrices}
P_{12} &= -\left[
\begin{array}{c|c}
A+B_2K & B_2 \\
\hline
C_1 +D_{12}K & D_{12}
\end{array}
\right]
\\
\nonumber
P_{21} & = \left[
\begin{array}{c|c}
A+LC_2 & B_1+LD_{21} \\
\hline
C_2 & D_{21}
\end{array}
\right].
\end{align} 
\fi

Note that since $\mathcal{Y}\subset \RR\mathcal{X}$, $\hat{Y}$
is strictly proper, and $\hat{M}$, $\tilde{M}$ have identity
feed-through terms, the constraint in \eqref{youlaFIR} implies that
$V\in\RR\mathcal{X}$. 

For a fixed $V\in \RR\mathcal{X}$, the optimal $U\in \frac{1}{z^{N+1}}\RR\Htwo$ is found by solving 
\[
\min_{U\in \frac{1}{z^{N+1}} \RR\Htwo} \|P_{11}+P_{12}VP_{21}+P_{12}UP_{21}\|^2_{\Htwo}.
\]
A necessary condition for $U$ to be optimal, given $V$, is 
\[
P_{12}^{\sim}P_{11}P_{21}^{\sim}+P_{12}^{\sim}P_{12}VP_{21}P_{21}^{\sim}+
P_{12}^{\sim}P_{12}UP_{21}P_{21}^{\sim}\in \left(
\frac{1}{z^{N+1}}\Htwo
\right)^{\perp}.
\]
Lemma~\ref{lem:modelMatchProd} implies that 
$P_{12}^{\sim}P_{12} = \Omega$, $P_{21}P_{21}^{\sim} = \Psi$, and 
$\P_{\frac{1}{z}\Htwo}\left(
P_{12}^{\sim}P_{11}P_{21}^{\sim}
\right)=0$.
Thus, the optimality condition becomes 
\[
P_{12}^{\sim}P_{11}P_{21}^{\sim}+\Omega V\Psi+
\Omega U\Psi \in \left(
\frac{1}{z^{N+1}}\Htwo
\right)^{\perp}.
\]
Furthermore, $U$ must satisfy
\[
U= -\P_{\frac{1}{z^{N+1}}\Htwo}\left(
\Omega^{-1}P_{12}^{\sim}P_{11}P_{21}^{\sim}\Psi^{-1}+V
\right)=0.
\]
Thus, the optimal $U$ is $0$ for any $V\in\RR\mathcal{X}$. 

Plugging
$V$ into the cost of \eqref{eq:modelMatching} and applying Lemma~\ref{lem:modelMatchProd} 
gives 
\if\MODE1
\[
\|P_{11}+P_{12}VP_{21}\|_{\Htwo}^2 
= 
\|P_{11}\|_{\Htwo}^2 + \|P_{12}VP_{21}\|_{\Htwo}^2
+2\langle P_{11},P_{12}VP_{21}\rangle
 = \|P_{11}\|_{\Htwo}^2 + \|\Omega^{1/2} V \Psi^{1/2}\|_{\Htwo}^2.
\]
\fi
\if\MODE2
\begin{align*}
\MoveEqLeft
\|P_{11}+P_{12}VP_{21}\|_{\Htwo}^2 
\\
&
= 
\|P_{11}\|_{\Htwo}^2 + \|P_{12}VP_{21}\|_{\Htwo}^2
+2\langle P_{11},P_{12}VP_{21}\rangle
\\ &
 = \|P_{11}\|_{\Htwo}^2 + \|\Omega^{1/2} V \Psi^{1/2}\|_{\Htwo}^2.
\end{align*}
\fi

Thus, Theorem~\ref{thmYoulaDecentralized} and
the model matching formulation, \eqref{eq:modelMatching}, imply that
the optimal $V$ must solve \eqref{simplifiedModelMatching}. Note that 
\[
\|\Omega^{1/2}V\Psi^{1/2}\|_{\Htwo}^2 = \sum_{i=1}^N \Tr(\Omega V_i \Psi V_i^\tp)
\]
is a positive definite quadratic function of $V$, while the constraint
is linear. Thus \eqref{simplifiedModelMatching} is a quadratic program
and it must have a unique optimal solution. 
\end{IEEEproof}

\gap


For completeness, a state-space realization will be given
for $\K$ of the form 
$
\K = (\hat Y -
\hat M V)(\hat X - \hat N V)^{-1}
$
with $V\in \RR\mathcal{X}$. Note that $V$ has a realization 
\if\MODE1
\begin{equation*}
V = \left[
\begin{array}{cccc|c}
0_{q_2\times q_2 } & & & & I_{q_2} \\
I_{q_2} & 0_{q_2\times q_2} &  & &  0 \\
& \ddots & \ddots && \vdots \\
&& I_{q_2} & 0_{q_2\times q_2} & 0_{q_2\times q_2} \\
\hline
V_1 & V_2 & \cdots & V_N & 0_{p_2\times q_2}
\end{array}
\right] 
=: 
\left[
\begin{array}{c|c}
A_V & B_V \\
\hline 
C_V & 0
\end{array}
\right].
\end{equation*}
\fi
\if\MODE2
\begin{align*}
V &= \left[
\begin{array}{cccc|c}
0_{q_2\times q_2 } & & & & I_{q_2} \\
I_{q_2} & 0_{q_2\times q_2} &  & &  0 \\
& \ddots & \ddots && \vdots \\
&& I_{q_2} & 0_{q_2\times q_2} & 0_{q_2\times q_2} \\
\hline
V_1 & V_2 & \cdots & V_N & 0_{p_2\times q_2}
\end{array}
\right] 
\\ &
=: 
\left[
\begin{array}{c|c}
A_V & B_V \\
\hline 
C_V & 0
\end{array}
\right].
\end{align*}
\fi
Standard state-space manipulations show that 
\begin{equation}\label{KRealization}
\K = \left[
\begin{array}{cc|c}
  A+B_2K+LC_2 & B_2 C_V & -L \\
  B_V C_2 & A_V & -B_V \\
  \hline
  K & C_V & 0
\end{array}
\right].
\end{equation}
Thus, $\K$ has a state-space realization of order $n+q_2N$. If $N$ is
the smallest integer such that a decomposition of the form
\eqref{eq:subspace} holds, then $\K$ must have entries in
$\frac{1}{z^{N+1}}\RRp$. In this case, any minimal
realization must have order at least $N$.  Thus, the order of the
realization in \eqref{KRealization} is
within a constant factor of the minimal realization order.

\subsection{Vectorization}

In this subsection, the quadratic program of 
Theorem~\ref{thm:modelMatchQP} will be cast as a finite-horizon
state-feedback problem using vectorization techniques. 
The vectorization approach is similar to method
used in \cite{feyzmahdaviandistributed2012}. 

First, by defining $R = \Psi \otimes \Omega $, the vectorized form of
the cost function becomes 
\begin{equation}\label{vectorizedCost}
\|\Omega^{1/2}V\Psi^{1/2}\|_{\Htwo}^2 = \sum_{i=1}^N 
\Tr(\Omega V_i \Psi V_i^{\tp} ) = \sum_{i=1}^N \vec(V_i)^{\tp} R \vec(V_i).
\end{equation}

Define the FIR transfer matrix $J = \P_{\mathcal{X}}((-\hat Y + \hat M V) \tilde M)$. 
If $\mathcal{Y}$ is defined by \eqref{eq:subspace}, then the model
matching constraint of \eqref{eq:modelMatching} is equivalent to $J_i
\in \mathcal{Y}_i$. The vectorized form of $J$ is computed from
\[
\vec(-\hat Y\tilde M + \hat M V \tilde M) = -\vec(\hat Y \tilde M) + 
(\tilde M^\tp \otimes \hat M)\vec(V).
\]
By Lemma~\ref{lem:vectorization} in the appendix, the terms of $J_i$ can be computed by the recursion 
\begin{align}
\label{vectorizedRecursion}
x_{i+1} &= A_v x_i + B_v \vec(V_i),\quad  
x_1 = \begin{bmatrix}
\vec(L) \\ 0_{np_2\times 1}
\end{bmatrix}
\\
\nonumber
\vec(J_i) &= C_v x_i + \vec(V_i),
\end{align}
where
\if\MODE1
\begin{equation*}
\left[
\begin{array}{c|c}
A_v & B_v \\
\hline
C_v & D_v
\end{array}
\right]
=
\left[
\begin{array}{cc|c}
I_{q_2} \otimes (A+B_2K) & 0_{nq_{2}\times np_2} & I_{q_2} \otimes B_2 \\
C_2^\tp \otimes K & (A+LC_2)^\tp \otimes I_{p_2} & C_2^\tp \otimes I_{p_2} \\
\hline
I_{q_2} \otimes K & L^\tp \otimes I_{p_2} & I_{p_2 q_2}
\end{array}
\right].
\end{equation*}
\fi
\if\MODE2
\begin{align*}
\MoveEqLeft[1]
\left[
\begin{array}{c|c}
A_v & B_v \\
\hline
C_v & D_v
\end{array}
\right]
\\ &
=
\left[
\begin{array}{cc|c}
I_{q_2} \otimes (A+B_2K) & 0_{nq_{2}\times np_2} & I_{q_2} \otimes B_2 \\
C_2^\tp \otimes K & (A+LC_2)^\tp \otimes I_{p_2} & C_2^\tp \otimes I_{p_2} \\
\hline
I_{q_2} \otimes K & L^\tp \otimes I_{p_2} & I_{p_2 q_2}
\end{array}
\right].
\end{align*}
\fi

Now let  $E_i$ and $F_i$ be matrices with columns that form
orthonormal
bases of $\vec(\mathcal{Y}_i)$ and $\vec(\mathcal{Y}_i^{\perp})$,
respectively. The term $\vec(V_i)$ can then be decomposed as 
\[
\vec(V_i) = E_iu_i + F_iu_i^{\perp},
\]
for some vectors $u_i$ and $u_i^{\perp}$. 

Using \eqref{vectorizedRecursion}, the constraint that $J_i\in\mathcal{Y}_i$ can be
equivalently cast as 
\begin{equation}\label{decomposedConstraint}
F_i^{\tp} (C_vx_i+\vec(V_i)) = F_i^{\tp} C_v x_i +
u_i^{\perp}= 0. 
\end{equation}
Plugging \eqref{decomposedConstraint} into the cost
\eqref{vectorizedCost} and the recursion
\eqref{vectorizedRecursion} leads to the following optimal control
problem:
\begin{equation}\label{vectorizedControlProblem}
\begin{array}{cc}
\min_u & \sum_{i=1}^N 
\begin{bmatrix}x_i^\tp & u_i^\tp \end{bmatrix}
\begin{bmatrix}
C_i^\tp \\ D_i^\tp
\end{bmatrix}
\begin{bmatrix}
C_i & D_i
\end{bmatrix}
\begin{bmatrix}
x_i \\ u_i
\end{bmatrix} 
\\
\textrm{s.t.} &
x_{i+1} = A_i x_i + B_i u_i , \qquad
x_1 = \begin{bmatrix} \vec(L) \\ 0_{np_1\times 1}\end{bmatrix},
\end{array}
\end{equation}
where the time-varying matrices are given by\footnote{
This definition is a slight abuse of notation, since $B_i$ here are
distinct from the original input matrices $B_1$ and $B_2$, etc. Similarly for $C_i$ are distinct from the original output matrices.  
}
\begin{equation*}
\begin{bmatrix}
A_i & B_i \\
C_i & D_i
\end{bmatrix} = 
\begin{bmatrix}
A_v - B_v F_i F_i^\tp C_v & B_v E_i \\
-R^{1/2} F_iF_i^\tp C_v & R^{1/2} E_i
\end{bmatrix}.
\end{equation*}


As is standard, \cite{lewisoptimal1995}, the optimal controller can be
computed as $u_i = K_i x_i$, where
\[
K_i = -(R_i+B_i^\tp X_{i+1}B_i)^{-1}(B_i^\tp X_{i+1}A_i + D_i^\tp C_i)
\]
and $X_i$ is computed from backward recursion with $X_{N+1} = 0$ and 
\if\MODE1
\[
X_i = C_i^\tp C_i +A_i^\tp X_{i+1} A_i 
- (A_i^\tp X_{i+1}B_i +C_i^\tp D_i)(D_i^\tp D_i
+B_i^\tp X_{i+1}B_i)^{-1}(B_i^\tp X_{i+1} A_i +D_i^\tp C_i). 
\]
\fi
\if\MODE2
\begin{multline*}
X_i = Q_i +A_i^\tp X_{i+1} A_i 
\\ 
- (A_i^\tp X_{i+1}B_i +S_i)(R_i
+B_i^\tp X_{i+1}B_i)^{-1}(B_i^\tp X_{i+1} A_i +S_i^\tp). 
\end{multline*}
\fi
Furthermore, the optimal cost is given by $x_1^\tp X_1 x_1$. The next theorem follows immediately from
Theorem~\ref{thm:modelMatchQP} and the preceding discussion.

\begin{theorem}
{\it 
The optimal $V$ is computed as 
\begin{align*}
x_{i+1} &= (A_i+B_iK_i) x_i,\qquad 
x_1 = \begin{bmatrix} \vec(L) \\ 0_{np_1\times 1} \end{bmatrix},
\\
\vec(V_i) &= (E_iK_i - F_i F_i^\tp C_v) x_i.
\end{align*}
Furthermore, the decentralized $\Htwo$ problem of
\eqref{eq:standardProb} has optimal value 
$
\|P_{11}\|^2 + x_1^\tp X_1 x_1. 
$
}
\end{theorem}


%% file: examples.tex
\section{Numerical Examples}
\label{sec:examples}

This section gives some numerical examples of optimal controllers
computed using the vectorization method of the previous
section.\footnote{
Code for these examples is available at \url{http://www.ece.umn.edu/~alampers/code/decH2.php}
}


\subsection{The Chain Problem}


Recall the three-player chain structure from Figure \ref{fig:chain} with constraint specified by \eqref{eq:chain}.
Consider the plant with 
\if\MODE1
\[
A=\begin{bmatrix}
1.5 & 1 & 0 \\
1 & 1.5 & 1 \\
0 & 1 & 1.5
\end{bmatrix},
\quad
B=C^\tp = 
\left[
\begin{array}{cc:c}
I_{3\times 3} & 0_{3\times 3} & I_{3\times 3}
\end{array}
\right],\quad
D_{21} = D_{12}^\tp = 
\begin{bmatrix}
0_{3\times 3} & I_{3\times 3}
\end{bmatrix}
\]
\fi
\if\MODE2
\begin{gather*}
A=\begin{bmatrix}
1.5 & 1 & 0 \\
1 & 1.5 & 1 \\
0 & 1 & 1.5
\end{bmatrix},
\quad
B=C^\tp = 
\left[
\begin{array}{cc:c}
I_{3\times 3} & 0_{3\times 3} & I_{3\times 3}
\end{array}
\right],
\\
D_{21} = D_{12}^\tp = 
\begin{bmatrix}
0_{3\times 3} & I_{3\times 3}
\end{bmatrix}
\end{gather*}
\fi


For comparison purposes, the
optimal $\mathcal{H}_2$ norm was computed using model matching from
this paper and the LMI method of 
\cite{rantzerseparation2006,gattamigeneralized2006}. In both cases the norm
was found to be $34.9304$. In contrast, the centralized
controller gives a norm of  $24.236$.

\subsection{Increasing Delays}

\begin{figure}
\centering
\if\MODE1
\includegraphics[width=.5\columnwidth]{increasingDelayExp}
\fi
\if\MODE2
\includegraphics[width=.95\columnwidth]{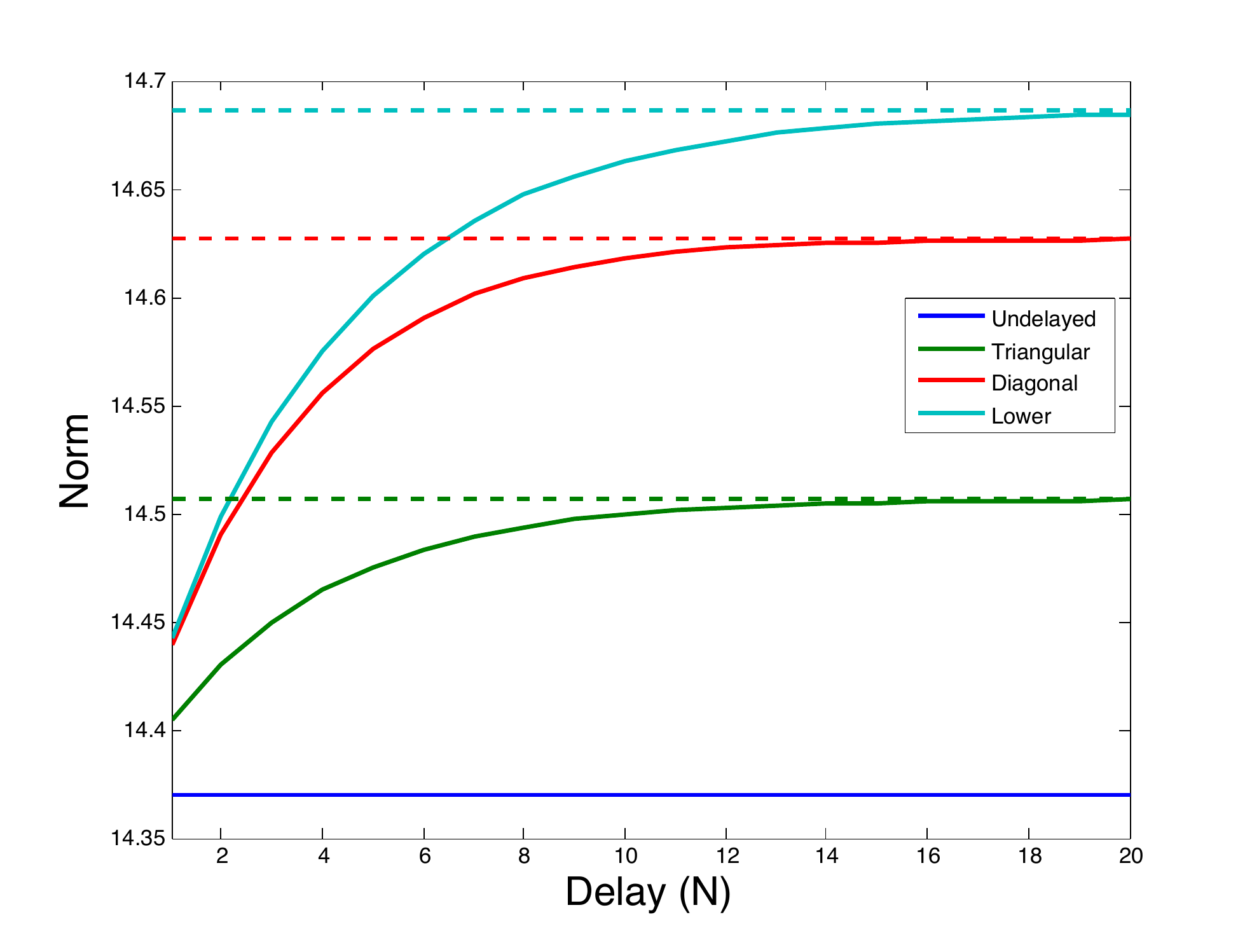}
\fi
\caption{\label{fig:exp} This plot shows the closed-loop norm for
 $\K_{\mathrm{Tri}}^N$, $\K_{\mathrm{Di}}^N$, and $\K_{\mathrm{Low}}^N$. For a given $N$, the controllers with
  fewer sparsity constraints give rise to lower norms.  
As $N$ increases, all of the norms
  increase monotonically since the controllers have access to less
  information. The dotted lines correspond to the optimal norms for
  sparsity structures given in \eqref{eq:sparseController}. 
}
\end{figure}

Consider the plant defined by 
\[
G = 
\left[
\begin{array}{c|cc:c}
\begin{bmatrix}
0.9 & 0 \\
0 & 1.1
\end{bmatrix}
& 1_{2\times 1} & 0_{2\times 2} & 0.1 I_{2\times 2} \\
\hline
1_{1\times 2} & 0_{1\times 1} & 0_{1\times 2} & 0_{1\times 2} \\
0_{2\times 2} & 0_{2\times 2} & 0_{2\times 2} & I_{2\times 2} \\
\hdashline
.1I_{2\times 2} & 0_{2\times 1} & I_{2\times 2} & 0_{2\times 2}
\end{array}
\right],
\]
where $1_{p\times q}$ is the $p\times q$ matrix of ones.

For $N\ge 1$, let $\K_{\mathrm{Tri}}^N$, $\K_{\mathrm{Di}}^N$, and $\K_{\mathrm{Low}}^N$ solve the decentralized $\Htwo$ problem, \eqref{eq:standardProb}, with constraints defined by
\if\MODE1
\[
\mathcal{Y}_{\mathrm{Tri}}^N = \bigoplus_{i=1}^N\frac{1}{z^i}
\begin{bmatrix}
\R & 0 \\
\R & \R
\end{bmatrix},
\quad
\mathcal{Y}_{\mathrm{Di}}^N = \bigoplus_{i=1}^N\frac{1}{z^i}
\begin{bmatrix}
\R & 0 \\
0 & \R
\end{bmatrix},
\quad
\mathcal{Y}_{\mathrm{Low}}^N = 
\bigoplus_{i=1}^N\frac{1}{z^i}
\begin{bmatrix}
0 & 0 \\
0 & \R
\end{bmatrix}.
\]
\fi
\if\MODE2
\[
\begin{array}{c}
\mathcal{Y}_{\mathrm{Tri}}^N = \bigoplus_{i=1}^N\frac{1}{z^i}
\begin{bmatrix}
\R & 0 \\
\R & \R
\end{bmatrix},
\quad
\mathcal{Y}_{\mathrm{Di}}^N = \bigoplus_{i=1}^N\frac{1}{z^i}
\begin{bmatrix}
\R & 0 \\
0 & \R
\end{bmatrix},
\\
\mathcal{Y}_{\mathrm{Low}}^N = 
\bigoplus_{i=1}^N\frac{1}{z^i}
\begin{bmatrix}
0 & 0 \\
0 & \R
\end{bmatrix}.
\end{array}
\]
\fi
The resulting norms are plotted in Figure \ref{fig:exp}.

In this example, as $N$ increases, the norms approach the optimal values given by the sparse controllers:
\if\MODE1
\begin{equation}
\label{eq:sparseController}
\K_{\mathrm{Tri}}^{s} \in \begin{bmatrix}
\frac{1}{z}\RRp & 0 \\
\frac{1}{z}\RRp & \frac{1}{z}\RRp
\end{bmatrix}
,\quad
\K_{\mathrm{Di}}^{s} \in 
\begin{bmatrix}
\frac{1}{z}\RRp & 0 \\
0 & \frac{1}{z}\RRp
\end{bmatrix}
,\quad
\K_{\mathrm{Low}}^{s} \in  
\begin{bmatrix}
0 & 0 \\
0 & \frac{1}{z}\RRp
\end{bmatrix},
\end{equation}
\fi
\if\MODE2
\begin{equation}
\label{eq:sparseController}
\begin{array}{c}
\K_{\mathrm{Tri}}^{\infty} \in \begin{bmatrix}
\frac{1}{z}\RRp & 0 \\
\frac{1}{z}\RRp & \frac{1}{z}\RRp
\end{bmatrix}
,\quad
\K_{\mathrm{Di}}^{\infty} \in  
\begin{bmatrix}
\frac{1}{z}\RRp & 0 \\
0 & \frac{1}{z}\RRp
\end{bmatrix}
\\
\K_{\mathrm{Low}}^{\infty} \in  
\begin{bmatrix}
0 & 0 \\
0 & \frac{1}{z}\RRp
\end{bmatrix},
\end{array}
\end{equation}
\fi
computed by the vectorization technique from
\cite{rotkowitzcharacterization2006}. Formally comparing the delayed solution to the sparse solution would be an interesting direction for future work. 


%% file: conclusion.tex
\section{Conclusion}
\label{sec:conclusion}

This paper derives a novel solution for a class of output feedback
$\Htwo$ control problems with quadratically invariant communication delay
patterns. First, all 
stabilizing decentralized controllers are characterized via doubly-coprime factorization. Then, by a
standard change of variables, the $\Htwo$ problem is cast as a convex
model matching problem. The main theorem shows that for a doubly-coprime
factorization based on the LQR and Kalman filter gains, the model
matching problem reduces to a quadratic
program. 
A solution to the quadratic program based on vectorization is also presented.  

The work in this paper has already been extended in
\cite{matnicommunication2014}, which builds on
\cite{matnidual2013,matnicommunication2013}, and uses the quadratic program 
in order to design communication structures for control via convex optimization.



Many open problems remain for optimal control with quadratically invariant
delay constraints. One limitation of the current work is that 
the solution is less explicit
than those available for state feedback
\cite{lamperskidynamic2012,lamperskioptimal2012} or special cases of
output feedback
\cite{sandellsolution1974,kurtaranlinearquadraticgaussian1974,yoshikawadynamic1975,feyzmahdaviandistributed2012,kristalnydecentralized2012}. 
It is hoped that the method in this paper can be utilized to derive
more general explicit solutions. Furthermore, work is needed to
understand how the controllers in this paper could be realized
\cite{vamsioptimal2011,lessardstructured2013} and
computed \cite{rantzerdynamic2009}  in a distributed
fashion.

%% file: acknowledgement.tex
\section{Acknowledgements}

The first author was supported by a Whitaker International
Postdoctoral scholarship. He would like to thank Nikolai Matni,
Michael Rotkowitz, and Laurent Lessard for helpful discussions. 


%% file: appendix.tex
\appendix

\renewcommand\thelemma{A.\arabic{lemma}}
\setcounter{lemma}{0}

This appendix collects state-space formulas that are useful for
deriving the results in the paper. For compactness, the proofs are
omitted or sketched. 


\begin{lemma}\label{lem:Prod}
{\it
Let $G$ and $H$ be real rational transfer matrices given by
\[
G = \left[
\begin{array}{c|c}
A_G & B_G \\
\hline 
C_G & D_G
\end{array}
\right]
\quad\textrm{and}\quad
H=\left[
\begin{array}{c|c}
A_H & B_H \\
\hline
C_H & D_H
\end{array}
\right]
\]
such that $A_G$ and $A_H$ are stable matrices and $G^{\sim}H$ is well
defined. Let $\Gamma$ satisfy the following Lyapunov equation:
$
\Gamma = A_G^\tp\Gamma A_H+C_G^\tp C_H.
$
Then the following equation holds
\begin{multline*}
G^{\sim}H = 
\left[
\begin{array}{c|c}
A_H & B_H \\
\hline
B_G^\tp \Gamma A_H + D_G^\tp C_H & D_G^\tp D_H + B_G^\tp \Gamma B_H
\end{array}
\right] 
\\
+
\left[
\begin{array}{c|c}
A_G & B_G \\
\hline
 B_H^\tp \Gamma^\tp A_G + D_H^\tp C_G & 0
\end{array}
\right]^{\sim} .
\end{multline*}
}
\end{lemma}

The following lemma is proved using Lemma~\ref{lem:Prod} and its
conjugated version.


\begin{lemma}\label{lem:modelMatchProd}
{\it
Let $P_{11}$, $P_{12}$, and $P_{21}$ be defined as
in \eqref{modelMatchMatrices}. The following equations hold:
\[
P_{12}^{\sim}P_{12} = \Omega, \qquad
P_{21}P_{21}^{\sim} = \Psi, \qquad
\P_{\frac{1}{z}\Htwo}(P_{12}^{\sim}P_{11}P_{21}^{\sim}) = 0. 
\]
}
\end{lemma}

\begin{lemma}\label{lem:vectorization}
{\it
For $\hat Y$, $\hat M$, and $\tilde M$ defined as in \eqref{ssDCF}, the following equation holds.
\if\MODE1
\begin{equation*}
\begin{bmatrix}
-\vec(\hat Y\tilde M) &
\tilde M^\tp \otimes \hat M
\end{bmatrix}
 = 
\left[
\arraycolsep=1.4pt
\begin{array}{cc|cc}
I_{q_2} \otimes (A+B_2K) & 0_{nq_{2}\times np_2} & \vec(L) &
I_{q_2} \otimes B_2 \\
C_2^\tp \otimes K & (A+LC_2)^\tp \otimes I_{p_2} & 0_{np_2\times 1} &
C_2^\tp \otimes I_{p_2} \\
\hline
I_{q_2} \otimes K & L^\tp \otimes I_{p_2} & 0_{p_2q_2\times 1} &I_{p_2 q_2}
\end{array}
\right].
\end{equation*}
\fi
\if\MODE2 
\begin{gather*}
\begin{bmatrix}
-\vec(\hat Y\tilde M) &
\tilde M^\tp \otimes \hat M
\end{bmatrix}
 = \\
\left[
\arraycolsep=1.4pt
\begin{array}{cc|cc}
I_{q_2} \otimes (A+B_2K) & 0_{nq_{2}\times np_2} & \vec(L) &
I_{q_2} \otimes B_2 \\
C_2^\tp \otimes K & (A+LC_2)^\tp \otimes I_{p_2} & 0_{np_2\times 1} &
C_2^\tp \otimes I_{p_2} \\
\hline
I_{q_2} \otimes K & L^\tp \otimes I_{p_2} & 0_{p_2q_2\times 1} &I_{p_2 q_2}
\end{array}
\right].
\end{gather*}
\fi
}
\end{lemma}

\begin{IEEEproof}
For more compact notation, let $A_K = A+B_2K$ and $A_L = A+LC_2$. The Kronecker product $\tilde M^\tp \otimes \hat M$ is computed as:
\if\MODE1
\begin{equation*}
\tilde M^\tp \otimes \hat M = (\tilde M^\tp \otimes I_{p_2} ) 
(I_{q_2} \otimes  \hat M) 
= \left[
\arraycolsep=1.4pt
\begin{array}{c|c}
A_L^\tp \otimes I_{p_2} & C_2^\tp \otimes I_{p_2} \\
\hline 
L^\tp \otimes I_{p_2} & I_{q_2}\otimes I_{p_2}
\end{array} 
\right]
\left[
\begin{array}{c|c}
I_{q_2} \otimes A_K & I_{q_2} \otimes B_2 \\ 
\hline
I_{q_2} \otimes K & I_{q_2}\otimes I_{p_2}
\end{array}
\right].
\end{equation*}
\fi
\if\MODE2
\begin{align*}
\tilde M^\tp \otimes \hat M &= (\tilde M^\tp \otimes I_{p_2} ) 
(I_{q_2} \otimes  \hat M) \\
&= \left[
\arraycolsep=1.4pt
\begin{array}{c|c}
A_L^\tp \otimes I_{p_2} & C_2^\tp \otimes I_{p_2} \\
\hline 
L^\tp \otimes I_{p_2} & I_{q_2}\otimes I_{p_2}
\end{array} 
\right]
\left[
\begin{array}{c|c}
I_{q_2} \otimes A_K & I_{q_2} \otimes B_2 \\ 
\hline
I_{q_2} \otimes K & I_{q_2}\otimes I_{p_2}
\end{array}
\right].
\end{align*}
\fi
Computing $-\vec(\hat Y\tilde M)$ is similar after noting that 
$
-\vec(\hat Y\tilde M) = \left(\tilde M^\tp \otimes
\left[
\begin{array}{c|c}
A_K & I_n \\
\hline
K & 0_{p_2\times n}
\end{array}
\right]
\right) \vec(L).
$
\end{IEEEproof}
